# Optical coherence tomography imaging of evoked neural activity in sciatic nerve of rat

J. Hope[1,2], M. Goodwin[1,2], F. Vanholsbeeck[1,2]

[1]Dodd Walls Centre for Photonic and Quantum Technologies, Auckland, New Zealand
[2]The Department of Physics, The University of Auckland, Auckland, New Zealand

**Abstract**

**Significance:** Imaging neural activity in myelinated tissue using optical coherence tomography (OCT) creates new possibilities for functional imaging in the peripheral and central nervous systems.

**Aim:** To investigate changes in OCT images in response to evoked neural activity in sciatic nerve of rat *in vitro.*

**Approach:** M-scans were obtained on peripheral nerves of rat using a swept source polarisation sensitive OCT system, while a nerve cuff acquired electrical neural recordings. From a total of 10 subjects: 3 had no stimulation (controls), 3 had paw stimulation, and 4 had nerve stimulation. Changes in the OCT signal intensity, phase retardation, phase, and frequency spectra were calculated for each subject and reference samples of a mirror and microspheres in solution.

**Results:** Observed changes in intensity in 3 paw stimulation and 2 nerve stimulation subjects and changes in frequency spectra amplitude in 2 paw stimulation subjects were above the reference noise level and were temporally consistent with osmotic swelling from ion currents during neural activity.

**Conclusion**: Light scattering changes produced by osmotic swelling, which have previously been characterised in squid and crab nerve, are also thought to occur in myelinated fibres on a scale which is detectable using OCT.

1. Introduction

Real-time functional monitoring of neural activity is a sought-after goal for diagnosis and treatment of neurological disorders [1] as well as fundamental research in neuroscience [2]. Currently, neural imaging modalities exhibit a trade-off between spatial resolution, temporal resolution, and imaging depth. Of the existing optical modalities, optical microscopy techniques based on voltage sensitive dyes and calcium indicators have adequate spatiotemporal resolution but severely limited imaging depth [3], whereas diffuse optical tomography has adequate imaging depth but critically low spatiotemporal resolution [4]. In optical coherence tomography (OCT), the spatiotemporal resolution of 10's of μs and several μm, and the imaging depth of 100's of μm in scattering media places the technology well for emerging clinical and research applications in neurology such as the development of structurally accurate models in peripheral nerve modulation treatments [5], and whole-brain neural network connectivity studies in mouse brain [2].

In axons and neurons, the rapid changes in membrane potential during an action potential are generated by ion currents across the cell membrane through voltage-gated ion channels with fast kinetics. Following an action potential, concentration dependent processes with comparatively slower kinetics transport the ions back across the membrane via ion pumps or transport the ions away from the axon or neuron via diffusion and via ion channels on glial cell membranes [6, 7]. Ion transport leads to transport of water molecules which surround dissociated ions, creating osmotic swelling, with inward ion currents producing osmotic swelling in the neuron or axon and outward ion currents producing osmotic swelling in the extracellular compartments which exist between axons or neurons and the surrounding basement membrane, glial cells, and tissue matrix, Figs. 1a and 1b [8, 9]. The relationship between these ion transport processes and observed optical effects have been proposed by many researchers, including: birefringence changes from voltage-dependent molecular realignment [10] scattering changes from osmotic swelling of cells [11] and extracellular compartments [12], and phase changes from membrane displacement in response to osmotic swelling of cells [13, 14] or in response to voltage induced curvature strain on the cell membrane known as the flexo-electric effect [15, 16]. A recent study on the flexo-electric effect presented a finite element model of cellular shape changes driven by ion concentration and membrane tension and favourably compared the model predictions of membrane displacement to interferometric measurements on cultured neurons [16].

In OCT, an interferometer measures the time-of-flight of light from a low coherence source which is reflected by scatterers within a sample. Analysis of the reflected light can be used to determine the sample optical properties such as refractive index, optical axis, birefringence and scatterer size. Researchers have used these optical properties to distinguish neural tissue from vasculature, meninges, bone, and connective tissue [17-21] in static images with 2 and 3 spatial dimensions called B-scans and C-scans, respectively. Dynamic imaging of neural tissue during neural activity requires imaging of the associated temporal changes in optical properties. Slow changes caused by haemodynamics occur in the order of seconds and have been imaged in cortex using intravascular tracer [22], doppler [23], and spectroscopic [24] based OCT techniques which analyse the properties of several temporally spaced B-scans. While fast changes caused by electrical activity occur in the order of milliseconds, and have been imaged in abdominal ganglion of sea slug and in unmyelinated axons of squid and crayfish using intensity [11, 14], birefringence [25], and phase [13-15] based OCT techniques applied to images with one spatial (depth) and one temporal dimension, called an M-scan, or in one study [25] with temporally spaced B-scans. While the results of these studies [11, 13-15, 22-25] demonstrate great promise for OCT in functional imaging of neural activity, in several studies [11, 13-15, 25] the tissues were non-mammalian and so are not easily translatable to clinical applications for humans, and in all studies the tissues were unmyelinated and so differ in cytostructure from myelinated fibres, Fig. 1a and 1b. Expanding functional imaging with OCT from unmyelinated fibres and neurons to myelinated fibres is important because the latter have distinct roles in communication within the central and peripheral nervous systems, and unique pathologies such as multiple sclerosis [26].

Myelin is produced by oligodendrocyte and Schwann glial cells, which are ubiquitous in mammalian peripheral nerves and white matter of the brain and spinal cord and exhibits a strong birefringence [27]. In myelinated nerve, dynamic changes in the optical birefringence during neural activity have been characterised in a transmission polarisation study on sciatic nerve of mouse [28] revealing relative changes in the total acquired signal in the order of 0.1 %, whereas dynamic changes in intensity have been modelled for a time domain OCT system [29] which estimated depth resolved changes in the acquired signal in the order of $1\times10^{-4}$ %.

In the latter, visual inspection of predicted dynamic changes in an A scan also reveals possible changes in the phase and the frequency spectra from Mie scattering. Light scattering from osmotic swelling would be expected to exist in myelinated fibres where, during an action potential, the potassium ion currents cross the paranode axon membrane to the periaxonal space via voltage gated ion channels with fast kinetics, and then, following an action potential, concentration dependent processes with comparatively slower kinetics transport the potassium ions via ion pumps back across the axon membrane, or via diffusion through the myelin sheath and along the periaxonal space to the extracellular fluid [6, 8], Fig. 1b. The flexo-electric effect might also exist in myelinated fibres and generate membrane displacement in the sheathed axons, although models have not yet been extended to encompass myelin sheaths [16].

In the current study, we used M-scans recorded using a swept source polarisation sensitive optical coherence tomography (PS-OCT) system to investigate changes in intensity, phase retardation, phase, and frequency spectra in response to evoked neural activity acquired in sciatic nerve of rat *in vitro*. The changes in these parameters were compared to electrical neural recordings acquired using a nerve cuff implanted on the sciatic nerve adjacent to the OCT recording site. The temporal characteristics of observed changes in intensity and frequency spectra amplitude correlated well with underlying physiological activity, creating new possibilities for functional imaging in peripheral nerves and white matter of the brain.

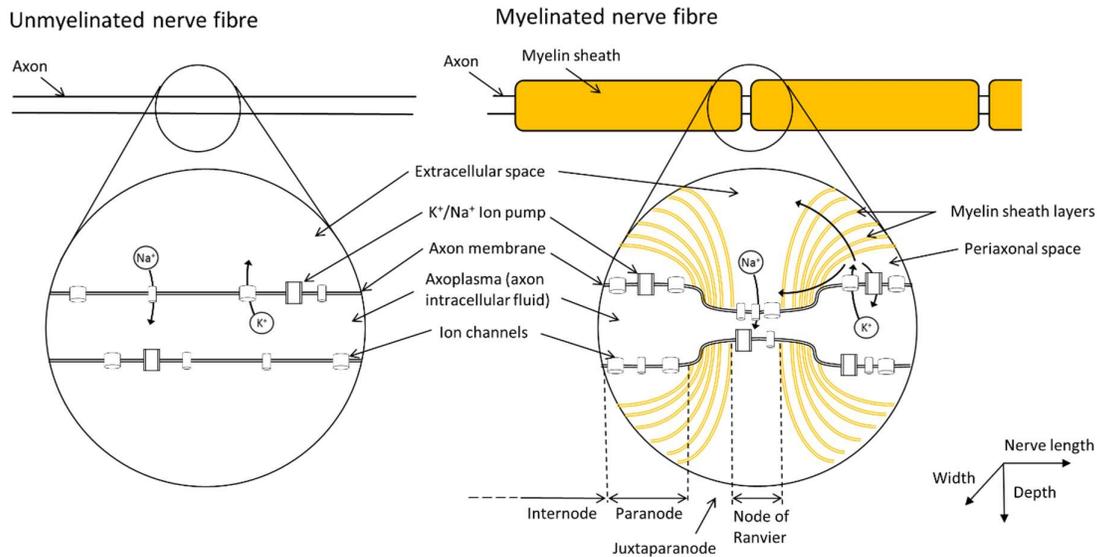

Figure 1: Schematics of unmyelinated (a) and myelinated (b) nerve fibres showing the ion current pathways across the axon membrane and, in the myelinated fibre, across the myelin sheath layers in the paranode (Sodium: Na+, Potassium: K+). The kinetics of the ion current pathway through the myelin sheath is comparatively slower than across the axon membrane causing accumulation of potassium ions in the periaxonal space and the myelin sheath layers, leading to swelling in these areas from osmosis. Not shown in the figure is the basal membrane which surrounds each fibre, separating it from the extracellular space and trapping a thin extracellular fluid layer around the fibre.

## 2. Methods

*2.1. Tissue preparation and handling*

Animal procedures were approved by the University of Auckland Animal Ethics Advisory Committee. A total of ten subjects were used in the current study, all Sprague Dawley rats, in the weight range 300 – 700 g, and of either gender. All subjects were used in *in-vitro* experiments after being euthanized using carbon dioxide followed by cervical dislocation and were provided as part of routine population culls. After culling, the cadaver of each subject was transported to a neighbouring building for hardware implantation and data acquisition. To access the sciatic nerve, subjects were placed in the supine position, the left hind leg was de-gloved of skin, and an incision was made down the posterior side of the leg to expose the sciatic, peroneal and tibial nerves. The left hind limb was fixed in place using a strap around the hock.

In the paw stimulation protocol, stimulation pins were placed through toes 1-2 and 4-5 of the left hind paw, Figs. 2a and 2b; the nerve cuff was implanted around the peroneal and tibial nerves; and, the sciatic nerve was lifted using the top side of curved tip tweezers onto the 3D printed nerve-holder which was aligned to the focal point of the OCT incident beam. In the nerve stimulation protocol, stimulus was applied via electrodes on either side of the 3D printed nerve-holder, Figs. 2c and 2d; the nerve cuff was implanted around the sciatic nerve; the peroneal and tibial nerves were lifted onto the 3D printed nerve-holder; and, the distal end of the peroneal and tibial nerves were clamped using a haemostat to stop the evoked neural activity from reaching muscles and inducing a movement artefact. In both the paw and nerve stimulation protocols, the tissue ground pin was placed through the paw in the contralateral hindlimb; and, exposed nerve tissue was perfused with physiological saline solution to hydrate the tissue in each experiment after hardware implantation and before data acquisition.

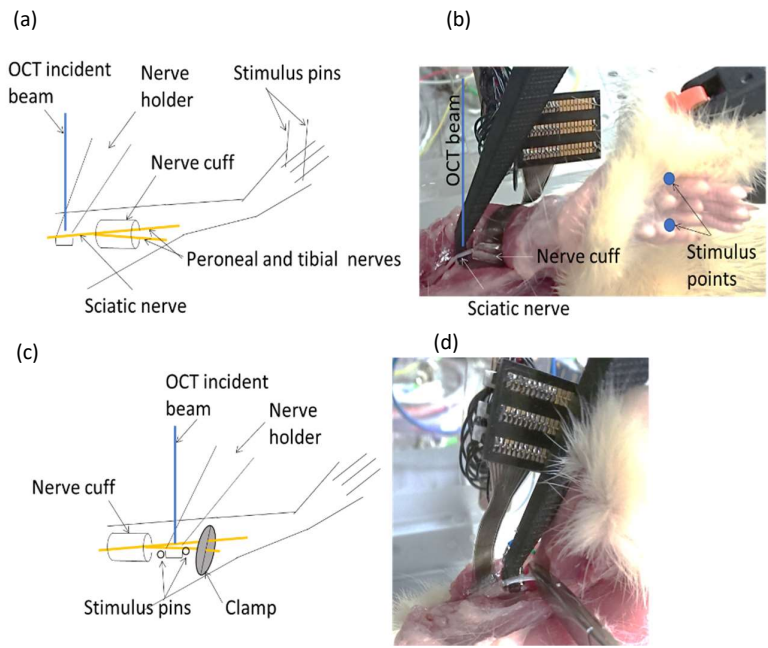

Figure 2: Schematic and picture of hardware interfacing the nerve in the paw stimulation (a – b) and the nerve stimulation (c – d) protocols.

## 2.2. Experiment apparatus

The PS-OCT system used in this study has previously been described in detail [30] and has an axial resolution of 10 μm in air and a lateral resolution of 20 μm. The system uses a swept source laser source (Axsun Technologies Inc, AXP50125-6) with a central wavelength of 1310 nm, bandwidth of 100 nm, and a 50 kHz sweep rate. A quarter wave plate in the sample arm is used to circularly polarise the incident light. The reflected light passes through the same waveplate before reaching the polarisation beam splitter, which split the light into its horizontal and vertical components, Fig. 2. The horizontal and vertical components then interfere with the light from their respective reference arms, the interference spectrum for both polarisation components are measured on two photodetectors (Thorlabs, PDB425C), and these measurements are digitized at 12 bits per sample and 125 MS/s (National Instruments, NI 5761). A digital trigger input line generated by an external waveform generator (Agilent, 33521A) initiates data acquisition and saving of an M-scan as .tdms file, where each M-scan contains either 5000 or 8000 A-scans temporally spaced at 20 μs (the laser sweep rate), and each A scan contains 1024 points. A galvanometer mirror (Thorlabs, GVS002) in the sample arm allows acquisition of B-scans but was switched off for M-scan acquisition to remove noise from mains and controller dither.

Stimulation pulses were generated by an isolated pulse stimulator (AM-systems, Model 2100) and were biphasic, anode-leading and +/-5 mA and 250 μs per phase in paw stimulation protocol, or +/-2 mA and 50 μs per phase in the nerve stimulation protocol. Each stimulation pulse was administered when triggered by a digital input line 16 ms after onset of the M-scan data acquisition, Fig. 3.

The nerve cuff electrode array was produced using the fabrication method described in Ref. [31]. The electrode array contained 2 rings of 14 electrodes, with each electrode 1.1 x 0.11 mm in area and spaced at a 26° pitch, and the two rings spaced 2 mm apart centre to centre. Electrodes were coated with poly(3,4-ethylenedioxythiophene):p-toluene sulfonate (PEDOT-pTS) to reduce the contact impedance, then assembled into a silicon nerve cuff using previously described methods [32]. The nerve cuff was soaked in phosphate buffered solution (PBS) between subjects and for one hour at the start of each day prior to being implanted. At the conclusion of experiments, the electrode array was cleaned with isopropyl alcohol and then stored in air.

Each of the two electrode rings in the nerve cuff were shorted together to produce two low impedance channels prior to connection to the headstage (INTAN, C3314) and then differential measurement was acquired between the two channels. Data were analogue band pass filtered (BPF) at 0.1 Hz to 5 kHz before being sampled at 30 kS/s, and then a 50 Hz notch filter was applied in software before data were saved to .rhd files. The digital line from the waveform generator which triggered OCT data acquisition was input to a digital input line on the USB interface (INTAN, C3100) to provide co-registration of OCT and electrophysiological data sets, Fig. 3.

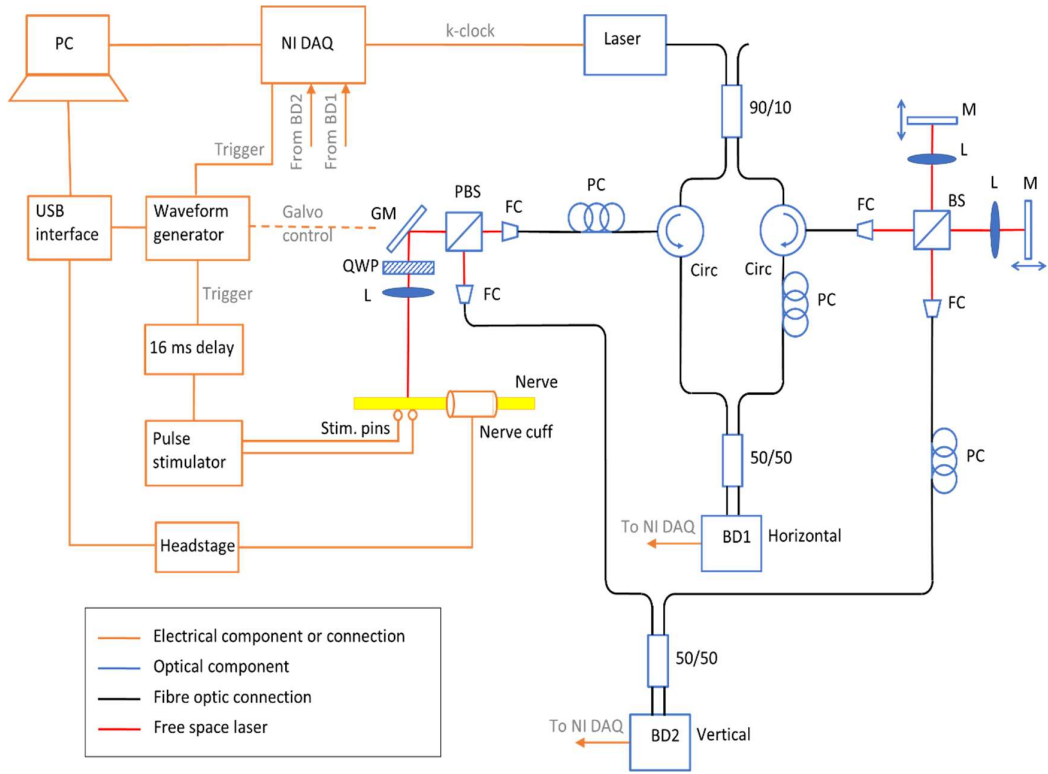

Figure 3: Schematic of the experiment apparatus configured for nerve stimulation, with the nerve shown in yellow. In paw stimulation the stimulus pins (Stim. pins) are inserted into the paw. PC: personal computer; NI DAQ: National instruments data acquisition card; GM: galvo mirror; QWP: quarter wave plate; L: lens; PBS: polarisation beam splitter; FC: fibre coupler; PC: polarisation controller; Circ: circulator; 50/50: 50/50 fibre coupler; 90/10: 90/10 fibre coupler; BS: 50/50 beam splitter; M: mirror on micro-translation stage; BD1 and BD2: balanced photodetector.

*2.3. Experiment protocols*

Four experiment protocols were performed, with one stimulus pulse administered per M-scan and 60 M-scans of OCT data collected from one nerve location for each subject. B-scans were acquired prior to commencement of each experiment to confirm the nerve position within the nerve holder and the incident beam position during M-scan acquisition.

i. System artefacts and noise were characterised by acquiring M-scan OCT data on a mirror with an absorptive filter in place to reduce reflected light to the nW range, and on a receptacle containing 2.5 wt% microspheres in aqueous solution with diameters of 3.8, 5.49 and 8.33 µm in equal parts.

ii. A control protocol was performed on three subjects to characterise artefacts introduced by neural tissue and physiological noise, where data were acquired from the OCT system and nerve cuff, but no stimulation pulses were administered.

iii. The paw stimulation protocol was performed on three subjects. Previous studies which characterised the compound action potential (CAP) produced using this stimulation technique in cadavers [32, 33] indicate excitation of myelinated somatosensory fibres, and, though not characterised, excitation of unmyelinated fibres was also expected.

iv. The nerve stimulation protocol was performed on four subjects. In the first of these four subjects, the peroneal and tibial nerves were not clamped to allow characterisation of the onset and duration of movement artefact. Direct stimulation of the nerve was expected to excite a larger number of nerve fibres and produce a larger CAP than paw stimulation based on previous studies [28, 34].

*2.4. Data processing*

### 2.4.1. Depth dimension processing

Four signal properties were analysed for neural activity based on existing studies: intensity [11, 29], phase retardation [14, 25, 28], phase [13, 15], and frequency spectra [29]. To determine these properties for each A-scan, raw data from each polarisation channel was Hanning windowed, zero-padded to $2^{14}$ points, then Fourier transformed to produce complex signals of the form:

$$F\{y_c\} = A_c e^{-i\phi_c} \qquad (1)$$

where $F\{y_c\}$ is the Fourier transform of raw data $y$, $A$ is the amplitude, $\phi$ the phase, and subscript $c$ indicates either the horizontal ($h$) or vertical ($v$) polarization channel. Zero-padding increased the axial sampling interval in air from 10 to 0.625 µm, and in neural tissue from 7.1 to 0.44 µm using a refractive index of 1.41 [20]. The square root of the intensity, $\sqrt{I}$, is equivalent to the overall amplitude and was calculated from the square root of the sum of squares of the horizontal and vertical channel amplitudes, Eqn. 2. The root intensity metric was used instead of intensity because the former provides improved contrast at greater depths beneath the nerve surface, though with the trade-off of lower signal to noise ratio. The phase retardation due to birefringence, $\theta$, was calculated from the angle between the two channel amplitudes, Eqn. 3, and the phase of the signal, $P$, was taken as the vertical channel phase, Eqn. 4.

$$\sqrt{I} = \sqrt{A_h{}^2 + A_v{}^2} \qquad (2)$$

$$\theta = \left(\frac{A_v}{A_h}\right) \qquad (3)$$

$$P = \phi_v \qquad (4)$$

Mie scattering theory relates the frequencies present in the backscattered light of an A-scan to the scatterer size and has been used previously in OCT studies on scatterer size analysis in biological samples [20, 35, 36]. In the current study, the frequency spectra, $S$, were calculated from a region of interest in the intensity of each A-scan which spanned from the maxima of the first peak, corresponding to the nerve surface, to a depth of 1000 pixels beneath the surface, corresponding to approximately 440 µm. This region of interest was mirrored around the surface peak, Gaussian windowed ($\alpha = 4$), zeropadded to $2^{16}$ points, and then Fourier transformed to obtain the complex spectra. The amplitude of the complex spectra was used for frequency spectra analysis. Zero padding data prior to the Fourier transform improves peak definition in the time frequency transformation by increasing the sampling density and has been employed previously in frequency analysis of time domain OCT data of neural activity [42].

### 2.4.2. Time dimension processing

The 2-D arrays of intensity, phase retardation, phase, and frequency spectra values were ensemble averaged across the 60 M-scans acquired from each subject to reduce noise, Fig. 4. Ensemble averaging is an element-wise process which reduces the amplitude of randomly occurring features but preserves features which are consistent between the averaged arrays and is a common noise reduction strategy in electrophysiology studies where a repeated, time-locked stimulus is available. Depth dependent baseline values were calculated as the temporal average across the first 10 ms of each array at each depth (or each frequency for spectra) which precedes the stimulus pulse and so was representative of the nerve absent of neural activity. These baseline values were then used to convert the M-scans of intensity, phase retardation, phase and frequency spectra into relative changes from baseline. Intensity and frequency spectra M-scans were scaled using scaling factors to account for variability in power of backscattered light caused by differences in OCT system calibration and nerve alignment across subjects. The scaling factors were calculated as the reciprocal of the sum of all values in the first A-scan of the M-scan and were normalised against the largest scaling factor. Percentage changes in the root intensity from baseline were also calculated but were not used for the initial feature identification because of its noise amplifying effect.

System artefacts and noise were characterised using M-scan data acquired on the mirror and on the microsphere sample. On the mirror data, system artefacts were identified as temporal changes in the depth position of the peak corresponding to the mirror surface. On microsphere sample data, system artefacts were visually identified as temporal changes in the 2D array of frequency spectra values. The system noise was sampled from the deepest 1000 pixels of the first A-scan in the intensity array and was analysed during ensemble averaging to characterise the noise reduction of this process. Noise was reduced using a low pass filter ($4^{th}$ order Butterworth, zero phase-shift) with 2 kHz cut off frequency after padding the arrays on either edge using mirrors of the arrays to avoid edge artefacts. Low frequency artefacts from system instability were removed using a high pass filter ($4^{th}$ order Butterworth, zero phase-shift) with 7.5 or 8 Hz cut off frequency, which was selected to have a period comparable to the unpadded array length, of 100 or 160 ms, and to be a divisor of the padded array length, of 300 or 480 ms (15,000 or 24,000 time increments).

Tissue artefacts were characterised using M-scan data acquired with the control protocol. Assuming the absence of system artefacts, a static tissue sample which is not undergoing any temporal changes in macro- or micro-structure shouldn't produce any temporal variation in intensity, phase retardation, phase, and frequency spectra. However, temporal changes in tissue temperature, health (due to lack of perfusion), and even mechanical relaxation of tissue, are all present and may affect the intensity, phase retardation, phase, and frequency spectra. Therefore, temporal changes in these parameters observed in the control protocol were assumed to be tissue artefacts and were used as a reference when analysing features in the paw and nerve stimulation subjects.

The combination of high and low pass filters, which created a BPF with passband of 8 Hz to 2 kHz, Fig. 4, was not expected to remove any features from neural activity based on frequency spectra of local field potentials and bioimpedance during neural activity observed in a previous study [32]. During data processing a Tukey window with an 80 % plateau was applied before the BPF.

The movement artefact produced using nerve stimulation protocol, as a result of excitation of motor fibres, was visually characterised on one subject. The nerve clamping method used in the remaining subjects induces a significant nerve crush injury and so prevention of action potentials propagating past this injury site can be considered highly unlikely. This was confirmed visually during experiments by the absence of twitching in the foot and hock synchronous with the stimulus.

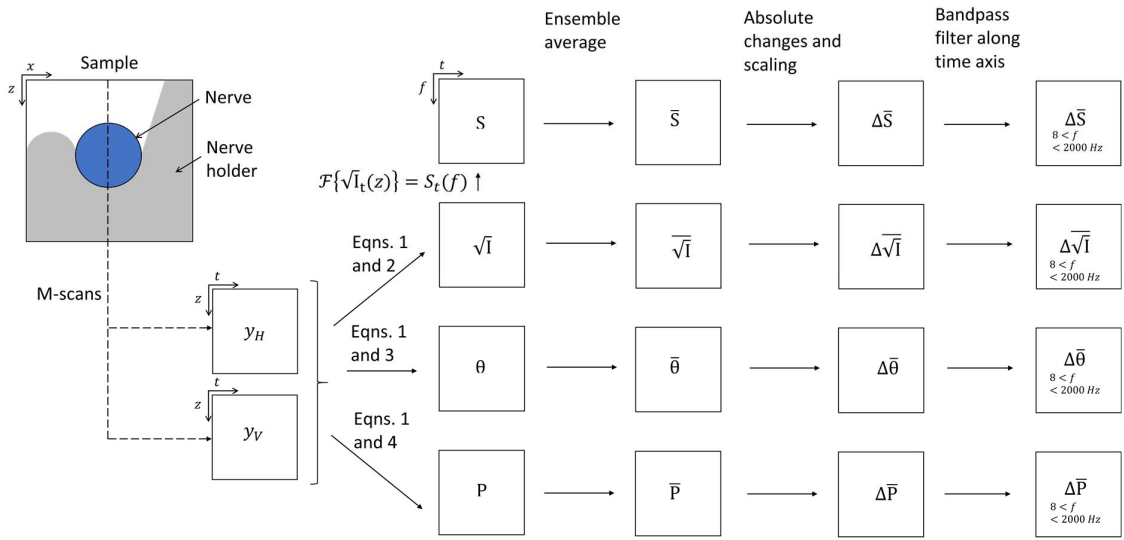

Figure 4: Data processing steps showing how the change in root intensity ($\Delta\overline{\sqrt{I}}$), phase retardation ($\Delta\overline{\theta}$), phase ($\Delta\overline{P}$), and spectra ($\Delta\overline{S}$), are calculated from raw M-scan data ($y_H$ and $y_V$) acquired on the PS-OCT system. Axes showing width ($x$), depth ($z$), time ($t$), and frequency ($f$) are only shown when they change from the previous step.

3. Results

*3.1. Mirror and microsphere sample*

Zero-padding raw A-scan data to $2^{14}$ points prior to equation 1 increased the axial sampling interval in air to 0.625 µm. On the mirror sample, the depth position of the peak corresponding to the surface of the mirror, varied by between 0 and 1 pixel, or 0 to 0.625 µm, across each M-scan and by 10 pixels, or 6.25 µm, across the 60 M-scans analysed. Ensemble averaging 60 M-scans reduced the standard deviation of the system noise by an order of magnitude. Further reduction in noise would be possible by ensemble averaging more data sets but with a lesser effect due to an exponential relationship between noise and number of data sets. For illustrations of the system stability and noise, refer to Fig. S1 in the Supplemental Materials.

On the mirror sample, Low pass filtering with a 2 kHz cut-off frequency reduced the noise by a factor of up to 5 but revealed low frequency changes, particularly in the intensity and phase arrays. High pass filtering with a 7.5 or 8 Hz cut-off frequency successfully removed these low frequency artefacts which were expected to be caused by forms of system instability such as drift in mechanical fixtures and optical components. On the microsphere sample, the change in intensity, phase retardation, and phase arrays contained features which were larger in depth, time, and amplitude than were observed in the mirror sample arrays. These larger features were expected to be caused by variations in the microsphere sample, such as movement of microspheres from Brownian motion, and highlight the sensitivities of the intensity, phase retardation, phase, and frequency spectra to small changes in the properties of highly scattering media. For illustrations of the mirror and microsphere results, refer to Fig. S2 in the Supplemental Materials.

*3.2. Tissue handling*

The time between culling and data acquisition was 6 +/- 1 minutes across all 10 subjects. This delay was expected to reduce the amplitude of observed neural activity but was within the tissue response lifetime observed in earlier experiments [37]. The nerve holder comfortably fit the nerve without any significant mechanical loading and allowed for variable nerve sizes and shapes across the subjects, Fig. 5 and Supplemental Fig. S3. The relatively broad weight range, of 400 – 700 g, required the height of the optical head to be adjusted between subjects using a micrometre stage because of variation, of up to 15 mm, in the distance between the sciatic nerve and the tray on which the subjects lay in the supine position.

Fouling of electrodes in the nerve cuff, which accumulated over successive experiments, caused excessive noise in some of the nerve cuff recordings rendering the recordings unusable. However, compound action potentials were confirmed to be present in recordings with low noise, and were assumed to be present with high confidence in the remainder of the experiments due to minor twitching of toes in response to paw stimulation and, prior to application of the nerve clamp, twitching of the paw and hock in response to nerve stimulation. Furthermore, the stimulation and recording techniques employed have been well established in pervious experiments by the same authors [32], as well as generally in literature.

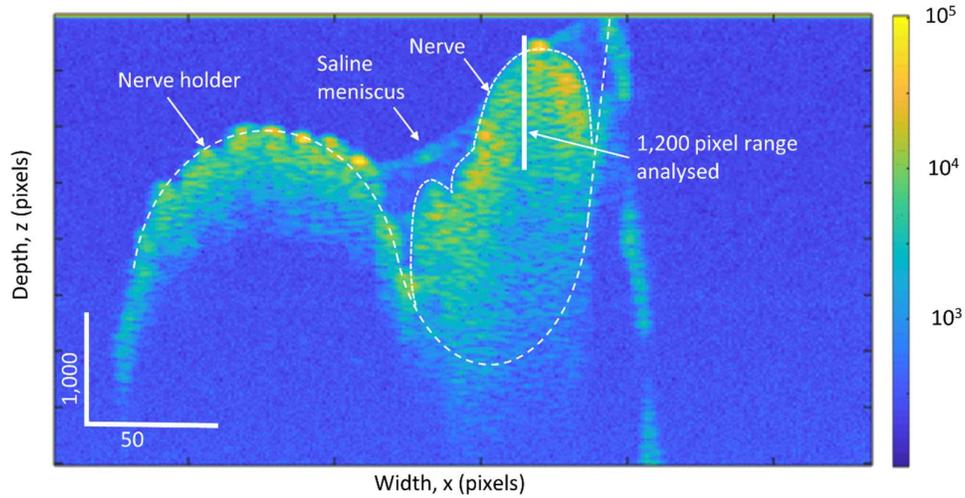

Figure 5: Labelled B-scan of a nerve in the nerve holder showing little distortion of the nerve from mechanical loading. Vertical and horizontal scale bars in all images are 1,000 and 50 pixels, respectively. Vertical pixel size is 0.44 µm in nerve tissue, Width pixel size is 20 µm. Colour axis is $\log_{10}$ of root intensity.

### 3.3. Control subjects

In all three control subjects, Figs. 6a to 6c, artefacts were present in the intensity, phase retardation, and phase arrays with amplitudes in the range of +/- 5 x $10^2$ a.u., +/- 4 x $10^{-2}$ rad, and +/- 3 x $10^{-1}$ rad, respectively, all of which were comparable to values observed in the filtered mirror sample (Fig. S2). These artefacts appeared visually similar to those observed in the microsphere sample (Fig. S2) albeit with significantly smaller amplitudes. In the control subjects, the amplitude of the changes in frequency spectra were a factor of 2 higher than the mirror sample. The tissue, therefore, did not introduce artefacts in the intensity, phase retardation, and phase arrays with amplitudes significantly above those caused by system noise and instability but did affect the frequency spectra. This difference in frequency spectra is likely caused by the absence of Mie regime scatterers in the mirror sample, making a poor reference for this comparison.

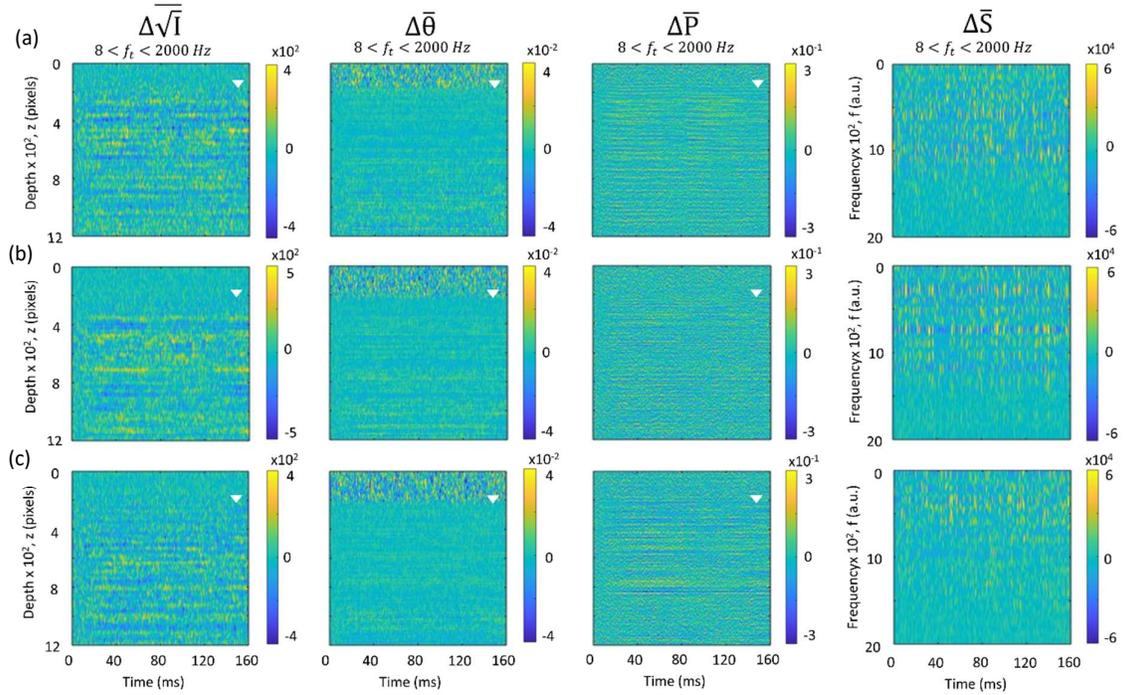

Figure 6: Change in intensity, phase retardation, phase, and frequency spectra arrays for control subjects (a – c). Bottom tips of the white triangles indicate the surface.

*3.4. Paw stimulation*

Nerve cuff recordings showed a compound action potential (CAP) which was 70 μV peak-peak amplitude, 2.5 ms in duration, and began approximately 2 ms after the stimulus pulse, Fig. 7a, which was in broad agreement with previous studies which used the same stimulus and recording technique [32, 33] albeit with a reduced amplitude due to the increased time between culling and data acquisition.

In all three paw stimulation subjects, a positive change in intensity with amplitudes of between 7 and 10 x $10^2$ a.u., amplitude peak at between 40 and 60 ms, and durations of approximately 40 ms, were visible in the intensity arrays, Figs. 7b to 7d. In the third subject, where the M-scan spanned 8,000 A-scans, a second positive change in intensity, with comparable amplitude and duration to the first peak, was observed in the intensity array at 110 ms, Fig. 7d. All intensity features were located between 40 and 150 pixels beneath the surface, which corresponds to approximately 18 to 68 μm in neural tissue [20] for the PS-OCT system. This depth, of 18 to 68 μm, correlates well to a previous study which characterised the thickness of the passive layer of epineurium tissue which covers the active neural tissue as approximately 45 +/- 30 μm [20], and is also significantly above the 10 pixel change in peak position from system instability (Fig. S1).

In the first and third of the paw stimulation subjects, positive changes in amplitude in the frequency spectra array of 16 x $10^4$ and 4 x $10^4$ a.u., respectively, were visible which temporally correlated with the observed changes in the intensity arrays, Figs. 7b and 7d. This latter amplitude, 4 x $10^4$ a.u., was below the noise level observed in the control subjects, Fig. 6. In all three subjects, features in the phase and phase retardation arrays were also visible which temporally correlated with the observed changes in the intensity arrays, although with amplitudes which were comparable to the noise levels observed in the control subjects.

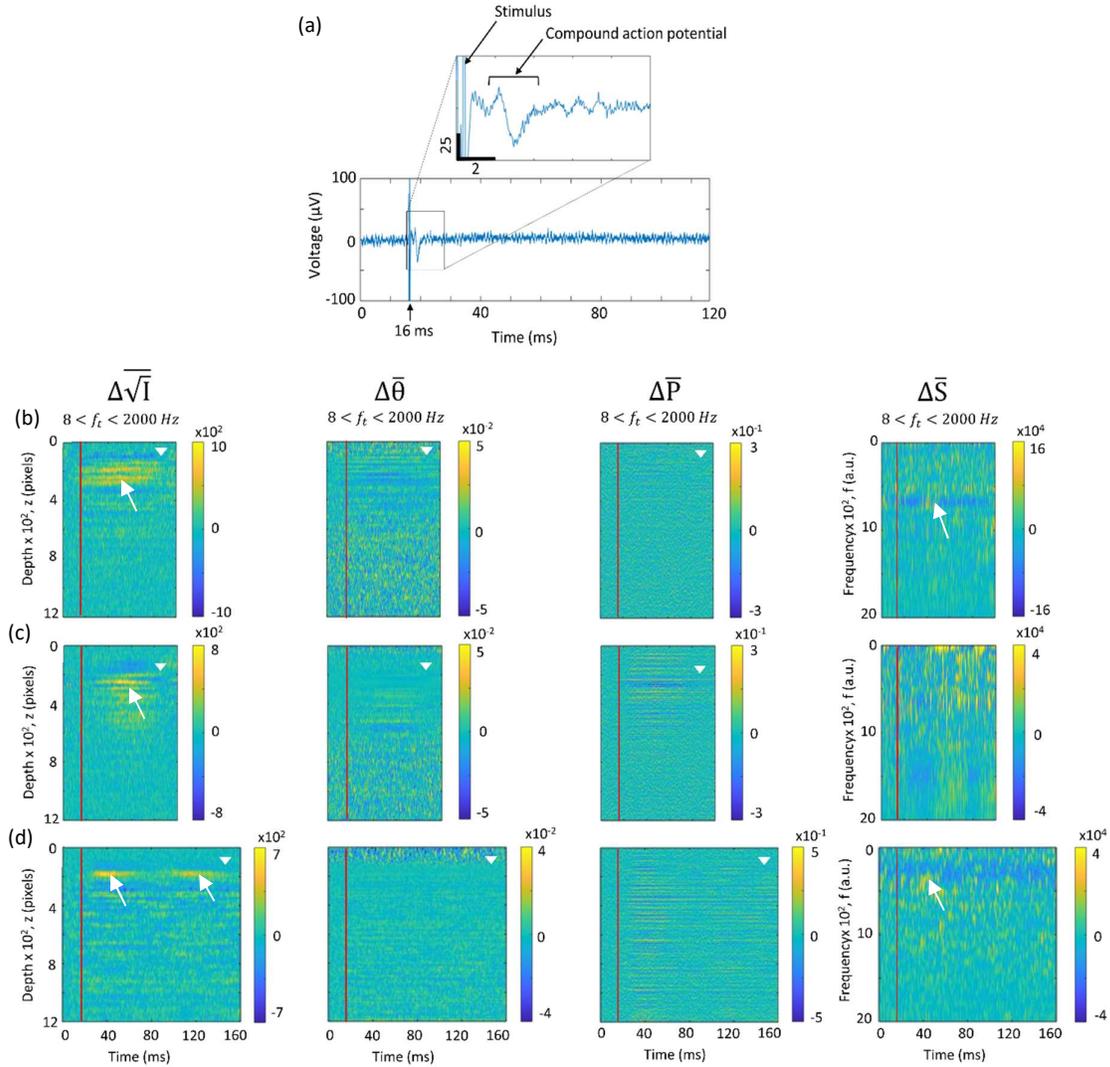

Figure 7: Nerve cuff recording of the compound action potential produced in the peroneal and tibial nerves in response to stimulation in the paw, with stimulation pulse at 16 ms (a). Change in intensity, phase retardation, phase, and frequency spectra arrays for paw stimulation subjects (b – d). Red vertical lines indicate when the stimulus was applied at 16 ms. White arrows identify positive changes in intensity thought to be neural activity produced in response to stimulus, and, where applicable, the corresponding positive changes in frequency spectra amplitude. Bottom tips of the white triangles indicate the surface.

*3.5. Nerve stimulation*

Nerve cuff recordings showed a compound action potential (CAP) which was 200 µV peak-peak amplitude, 4 ms in duration, and began less than 0.5 ms after the stimulus pulse, Fig. 8a. The larger amplitude than that of paw stimulation was expected because direct stimulation of the nerve excites more nerve fibres, whereas the longer duration was expected because the fast and slow neural activity, from myelinated and unmyelinated fibres, respectively, have not been separated by dispersion as is the case with paw stimulation.

In the first nerve stimulation subject, where the nerve was unclamped, stimulus was applied at 8 ms instead of 16 ms, and large changes in the intensity, phase retardation, phase and frequency spectra arrays were visible with amplitudes of, respectively, +/- 300 x $10^2$, +/- 40 x $10^{-2}$, +/- 4 x $10^{-1}$, and 160 x $10^4$ a.u, Fig. 8b. These amplitudes were significantly above those observed in the control and the paw stimulation protocols and were likely due to movement artefact produced by twitching of the hock and paw in response to excitation of motor fibres in the sciatic nerve. Further evidence for a movement artefact was the 5 to 10 ms delay between the stimulus and the onset of the positive change in the intensity and phase retardation arrays, which correlates well with the delay between neural action potential and muscle activation in a muscle twitch response called the latent period.

In the second nerve stimulation subject, Fig. 8b, the intensity, phase retardation, and phase arrays exhibited changes which were all within the amplitude range of the earlier control protocol arrays, Figs. 6a to 6c.

In the third and fourth nerve stimulation subjects, positive changes in intensity array were visible with amplitudes of 7 x $10^2$ to 10 x $10^2$, peaks at 40 to 60 ms, and commencing immediately following the stimulus, Figs. 8c and 8d. In both subjects, several further peaks, with comparable amplitudes and the same depths to the first peak, were visible at later times within the intensity arrays. As was the case with paw stimulation, the depth of the changes in intensity correlated well with the expected location of active neural tissue. However, unlike paw stimulation, with nerve stimulation the changes observed in the frequency spectra arrays did not clearly correlate to changes observed in the intensity arrays.

In all three nerve stimulation subjects with clamped nerves, features in the phase and phase retardation arrays were comparable in amplitude to the noise levels observed in the control subjects.

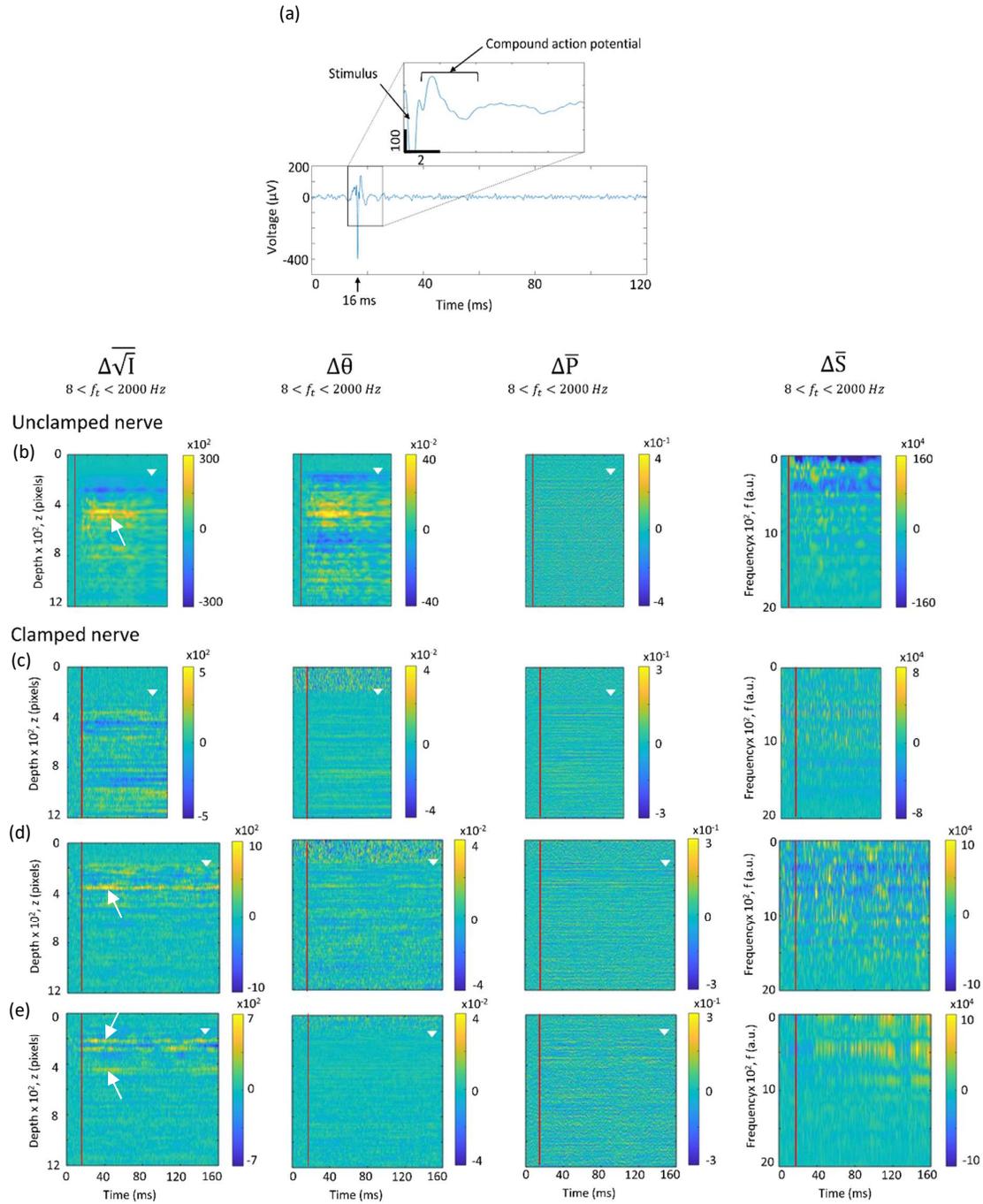

Figure 8: Nerve cuff recording of the compound action potential produced in the peroneal and tibial nerves in response to stimulation in the peroneal and tibial nerves, with stimulation pulse at 16 ms (a). Change in intensity, phase retardation, phase, and spectra arrays for nerve stimulation subjects with unclamped (b) and clamped (c – e) peroneal and tibial nerves. Red vertical lines indicate when the stimulus was applied at 8 ms (b) or 16 ms (c – e). White arrows identify positive changes in intensity thought to be caused by movement (b) or neural activity (d – e). Unlike with paw stimulation, in nerve stimulation there were no positive changes in frequency spectra amplitude which clearly corresponded to change in intensity. Bottom tips of the white triangles indicate the surface.

## 4. Discussion

Optical measurement of physiological changes in neural tissue during neural activity can be broadly classified as either targeting the changes caused by haemodynamics, which occur in the order of seconds [22-24], or the changes caused by ion transport, which occur in the order of milliseconds [11-16, 25, 28]. Of this latter group, most studies have measured fast changes in optical properties which temporally correlate with changes in transmembrane voltage during an action potential and the associated the transmembrane ion currents. Fewer studies have looked at the slower ion transport mechanisms associated with diffusion and osmotic swelling, and which manifest after the changes in transmembrane voltage [11, 12].

In the current study, the initial changes in intensity identified in paw and nerve stimulation subjects are thought to be caused by the slower ion current mechanism identified in unmyelinated axon of squid [12] where the observed change in light scattering had a time course of 100 to 200 ms and the action potential a time course of around 3 ms. Our assumption is supported by the prolonged duration, of 40 to 50 ms, of the observed changes in comparison to the 3 ms action potential, Fig. 9, which would be expected in myelinated nerve fibres based on recent modelling of potassium ion currents in myelinated fibres [6] and experimental observation in [8]. In the latter, sustained neural stimulation at 20 Hz elicited minimal morphological changes whereas 100 Hz elicited significant swelling in the paranode [8] thus placing the time course of some ion currents in the paranode at somewhere between 10 and 50 ms. A similar signal duration was observed in the intensity signal in OCT of neural activity in abdominal ganglion [11]. The amplitudes of the changes in intensity observed in paw stimulation and nerve stimulation subjects, of $7 \times 10^2$ to $10 \times 10^2$ a.u., correspond to relative changes in the range of 1 to 3 %, which are significantly larger than predicted by modelling in [29], of $1 \times 10^{-4}$ %. This difference may be due to the underlying mechanism, where in the latter, neural activity was modelled using the voltage dependent mechanism presented in [38], of a change in the refractive index of the axons related to the membrane potential, from light scattering measurements on unmyelinated axons. The negative changes in intensity which featured in paw and nerve stimulation subjects, often with a banding or alternating pattern with depth, for example in Figs. 7d and 8d, agreed with modelling in [29]. This agreement is likely because the voltage dependent and current dependent mechanisms both produce changes in the scattering properties in the Mie scattering regime.

In the paw stimulation subjects, an alternative possibility for the observed change in intensity and frequency spectra is movement artefacts since they share several features with the movement artefact produced in unclamped nerve stimulation: a positive change in intensity with a delay of several milliseconds and duration of several 10's of ms. Twitching in the paw in response to paw stimulation is limited to toes 2 to 4 (those between the stimulation pins) and contains in the order of 1 mm movement due to the lack of muscle tissue in the paw. While any resultant movement in the sciatic nerve is not visible, it is conceivable that the sensitivity of OCT is enough to detect micromotion. On the other hand, positive changes in intensity were observed in two of the three nerve stimulation subjects with clamped nerve, where all muscle activity was absent, and a movement artefact does not explain the presence of the second peak in the third paw stimulation subject, Fig. 7c. Instead, this second peak may be explained by unmyelinated neural activity, which is known to produce changes in intensity in squid axon in the range of 0.2 to 0.5 % [14]. At around 100 ms after the stimulus, the second peak temporally correlates with the expected time for unmyelinated neural activity to propagate from the paw to the sciatic nerve recording sites of 25 to 100 ms using a 0.5 to 2 m/s propagation velocity (as opposed to 30 to 120 m/s in myelinated fibres).

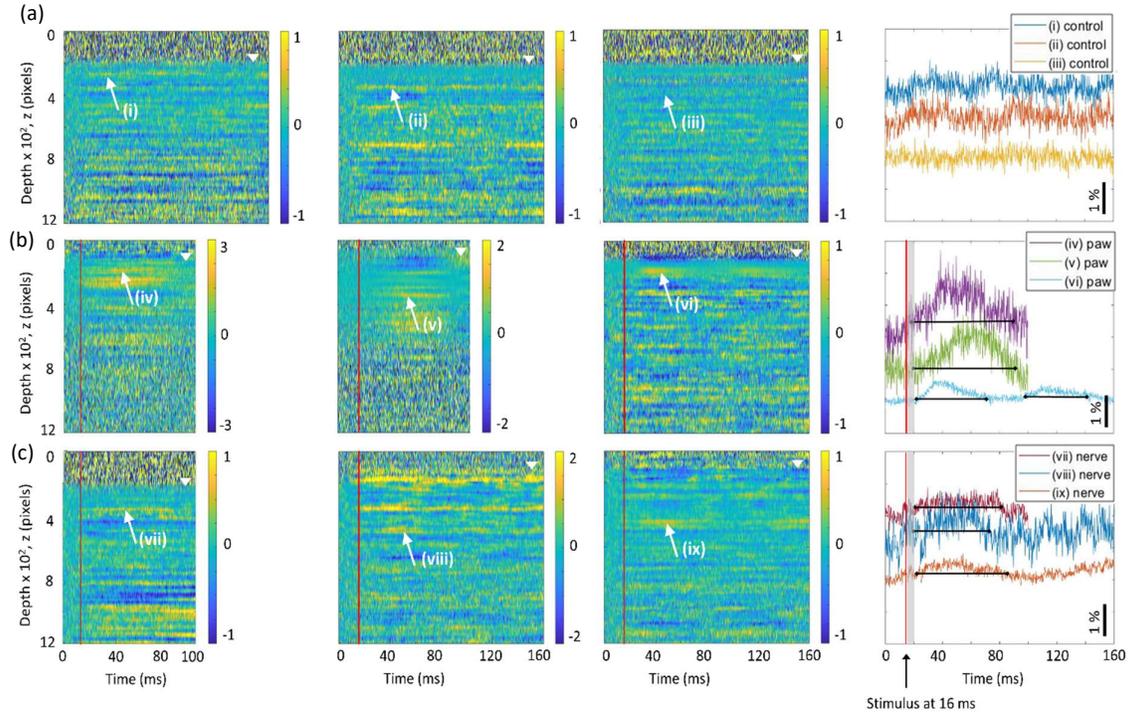

Figure 9: Relative change in intensity (%) for control (a), paw stimulation (b), and nerve stimulation (c) subjects. White arrows and roman numerals identify selected depths plotted on the right-hand side as time versus change in intensity (%). Bottom tips of the white triangles indicate the surface. Red vertical line indicates when the stimulus was applied. On the right-hand side, the grey shaded box indicates the 4 ms time period containing recorded myelinated neural activity, and the estimated onset and duration of features thought to be produced by neural activity are indicated by the overlaid black diamond-ended lines.

None of the paw stimulation and clamped nerve stimulation subjects produced features in the phase retardation array with amplitudes above those observed in the control subjects, of 0.3 rad. In an existing study which characterised changes in birefringence during neural activity in sciatic nerve of mouse using a transmission polarisation apparatus [28], the relative change was in the order of 0.1 %, which is an order of magnitude below the noise level of the PS-OCT system used in the current study; although, it should be noted, the transmission polarisation apparatus measured the change in intensity in one of the orthogonal polarisations while in the current study the phase retardation measured the change in angle created by the relative intensity in the two orthogonal polarisations, Eqn. 3. The higher sensitivity in transmission polarisation than in PS-OCT is likely due to the lower power of the light source in the PS-OCT system, of approximately 80 mW versus 700 mW, and additional noise in the PS-OCT system from combining measurement from two channels, Eqn. 3.

In all of the paw and nerve stimulation subjects, features were visible in the change in phase arrays which temporally correlate to changes in the intensity arrays, Figs. 7 and 8. However, these features were not analysed in the current study because their amplitudes, of 0.3 to 0.5 rad, were comparable to those of noise in the mirror sample and control subjects, of 0.3 rad. It is expected, however, that in a lower noise OCT system phase analysis would be able to detect changes in the neural tissue from osmotic swelling which was observed in the change in intensity arrays due to the sub resolution capabilities of phase analysis [39] and its previous success in detecting neural activity in unmyelinated fibres [13-15].

Changes in frequency spectra amplitude were not investigated in earlier neural OCT studies [11, 13-15, 25] and provides an interesting new avenue for exploration. In two of the paw stimulation subjects, the changes in frequency spectra amplitude which temporally correlated with the changes in intensity, Figs. 7b and 7d, were potentially both caused by the same osmotic swelling mechanism. In OCT, Mie scattering theory relates the frequency spectra of the signal in the depth dimension to the diameter of the scatterers and the refractive indices of the scatterers and surrounding media [35, 36]. The diameter range of axons in myelinated fibres, of around 1 to 10 μm, places them in the Mie scattering regime, whereas the thickness of the periaxonal space, of 0.1 μm, places it around the Rayleigh limit to Mie scattering and the thickness of myelin sheath layers, of less than 18 nm, places it well below the Rayleigh limit. Changes in volume from osmotic swelling in these three compartments – the axon, periaxonal space and myelin sheath layers – could significantly alter the Mie scattering and Rayleigh scattering contributions to the tissue scattering coefficient [40]. However, spectral changes which clearly correlated to the observed intensity changes were not observed in nerve stimulation subjects and further work is needed to understand the relationship between osmotic swelling and light scattering in myelinated fibres. In this regard, many of modelling components already exist, including models of myelinated fibre cytology and ion channel kinetics [6], models of osmotic swelling in response to ion currents in neural activity [9], and models of multilayer Mie scattering [41].

In the current study, the 50 kHz sweep rate of the swept source Laser provided sufficient temporal resolution for analysis of neural activity, while the PS-OCT system provided versatility by allowing calculation of intensity, phase retardation, phase, and frequency spectra from each set of horizontal and vertical channels M-scan data. Imaging of neural activity in peripheral nerves in-situ in cadavers allowed stimulation of peripheral tissue and nerve endings through the paw stimulation protocol, reduced the potential for nerve damage from tissue handling, and removed the need to ligate the nerve ends. On the other hand, imaging explanted nerve tissue in physiological solution, as was performed in previous studies [11, 13-15, 25], removes the possibility of movement artefacts from muscle activation. The nerve holder used in the current study maintained a stable nerve position under the optical head in the absence of movement artefacts from muscle activation, but also isolated the nerve from surrounding tissue and so likely accelerated cooling of the nerve. Continuous application of heated and oxygenated physiological saline to the nerve tissue during experiments, to stabilise nerve temperature and prolong tissue life, was decided against because a variable fluid layer on the tissue surface would affect the depth continuity of neural tissue in the M-scans. The sensitivity of intensity, phase retardation, phase and frequency spectra to movement artefacts, Fig. 8b, complicates translation of OCT techniques to in-vivo measurements of neural activity in nerves and likely necessitates use of muscle relaxants. For translation to in vivo measurements on white matter of the brain, traditional head fixing techniques might suffice.

In the control subjects, Figs. 6a to 6c, the consistency across subjects in the amplitude and visual appearance of artefacts in the intensity, phase retardation, and phase arrays provided good confidence that inactive tissue is optically stable over the course of the data acquisition time period. The comparable size in amplitude of artefacts in the control subjects to those in the mirror sample used to characterise the system noise level (Fig. S2) implied that any artefacts introduced by the tissue have negligible influence on the system noise and instability. Conversely, the amplitude of noise and artefacts in the frequency spectra array was significantly higher in the control subjects than in the mirror, and higher still in the microsphere sample, highlighting the sensitivity of the frequency spectra to noise and instability with highly scattering media.

Limitations of the current study are the use of cadavers instead of in-vivo subjects which limited the time for data collection and introduced a gradual degradation in tissue health due to lack of natural perfusion; the lack of temperature control in the neural tissue which slows the biokinetics of action potentials and introduces a non-natural tissue condition; the limited number of subjects in each protocol; and, drift and high noise in the OCT system. To address these limitations in future experiments, a portable OCT system is planned which can be assembled in the animal facility which will allow in-vivo experiments; and hardware modifications to the light source and galvanometer mirror are planned. To investigate a correlation between myelin swelling and OCT signal intensity, a future study would be useful which combines the OCT imaging and data processing techniques used in the current study with the sustained stimulation protocol in [8] and potassium ion current modelling in [6]. However, the long time-course of accumulative osmotic swelling in response to sustained stimulation, of several 10's of seconds to minutes, places a demanding requirement on OCT system stability.

5. **Conclusion**

Changes in root intensity, phase retardation, phase, and frequency spectra arrays were calculated from M-scans obtained using a PS-OCT system on peripheral nerves of rat in response to evoked stimulation in the paw and the nerve, as well as a control protocol with no stimulation. The temporal characteristics of observed changes in intensity and frequency spectra are thought to be caused by osmotic swelling from potassium ion currents in the nerve fibres. The sensitivity of OCT to osmotic swelling in myelinated fibres creates new possibilities for functional imaging in peripheral nerves and white matter of the brain.

**Data**

The data that support the findings of this study are openly available in Mendeley Data, V1: doi:10.17632/dh5z2hbzwd.1, doi:10.17632/2zwph29zdx.1, and doi:10.17632/w6z3g43syk.1.

**Disclosures**

The authors have no relevant financial interests in this article and no potential conflicts of interest to disclose.

**Acknowledgements**

The authors would like to acknowledge funding from Marsden Fund and Royal Society of New Zealand (UoA1509) which made this research possible; David Holder and Kirill Aristovich, from the EIT Research Group at University College London, for providing nerve cuff electrode arrays; Darren Svirskis and Mahima Bansal from the School of Pharmacy, The University of Auckland for their help with PEDOT coating electrodes; and staff at the School of Biological Sciences, The University of Auckland for their help with animal handling.

**References**

1. Morrison, R.A., et al., *Vagus nerve stimulation intensity influences motor cortex plasticity.* Brain Stimulation, 2019. **12**(2): p. 256-262.
2. Farahani, F.V., W. Karwowski, and N.R. Lighthall, *Application of Graph Theory for Identifying Connectivity Patterns in Human Brain Networks: A Systematic Review.* Frontiers in Neuroscience, 2019. **13**(585).
3. Chen, Y., et al., *Review of advanced imaging techniques.* Journal of pathology informatics, 2012. **3**: p. 22-22.
4. Dai, X., et al., *Fast noninvasive functional diffuse optical tomography for brain imaging.* Journal of Biophotonics, 2018. **11**(3): p. e201600267.
5. Grinberg, Y., et al., *Fascicular perineurium thickness, size, and position affect model predictions of neural excitation.* IEEE Transactions on Neural Systems and Rehabilitation Engineering, 2008. **16**(6): p. 572-581.
6. Brazhe, A.R., et al., *Excitation block in a nerve fibre model owing to potassium-dependent changes in myelin resistance.* Interface Focus, 2011. **1**(1): p. 86-100.
7. Jana, M. and K. Pahan, *Astrocytes, Oligodendrocytes and Schwann Cells*, in *Neuroimmune Pharmacology*. 2017, Springer. p. 117-140.
8. Trigo, D. and K.J. Smith, *Axonal morphological changes following impulse activity in mouse peripheral nerve in vivo: the return pathway for sodium ions.* The Journal of physiology, 2015. **593**(4): p. 987-1002.
9. Lee, J., D.A. Boas, and S.J. Kim, *Multiphysics Neuron Model for Cellular Volume Dynamics.* IEEE Transactions on Biomedical Engineering, 2011. **58**(10): p. 3000-3003.
10. Yao, X.-C., et al., *Cross-polarized reflected light measurement of fast optical responses associated with neural activation.* Biophysical journal, 2005. **88**(6): p. 4170-4177.
11. Graf, B.W., et al., *Detecting intrinsic scattering changes correlated to neuron action potentials using optical coherence imaging.* Optics express, 2009. **17**(16): p. 13447-13457.
12. Cohen, L.B., R.D. Keynes, and D. Landowne, *Changes in axon light scattering that accompany the action potential: current-dependent components.* The Journal of Physiology, 1972. **224**(3): p. 727-752.
13. Akkin, T., C. Joo, and J.F. de Boer, *Depth-Resolved Measurement of Transient Structural Changes during Action Potential Propagation.* Biophysical Journal, 2007. **93**(4): p. 1347-1353.
14. Akkin, T., D. Landowne, and A. Sivaprakasam, *Detection of Neural Action Potentials Using Optical Coherence Tomography: Intensity and Phase Measurements with and without Dyes.* Frontiers in neuroenergetics, 2010. **2**: p. 22.
15. Akkin, T., D. Landowne, and A. Sivaprakasam, *Optical Coherence Tomography Phase Measurement of Transient Changes in Squid Giant Axons During Activity.* Journal of Membrane Biology, 2009. **231**(1): p. 35-46.
16. Ling, T., et al., *High-speed interferometric imaging reveals dynamics of neuronal deformation during the action potential.* Proceedings of the National Academy of Sciences, 2020. **117**(19): p. 10278.
17. Islam, M.S., et al., *Extracting structural features of rat sciatic nerve using polarization-sensitive spectral domain optical coherence tomography.* Journal of Biomedical Optics, 2012. **17**(5).
18. Chen, Y., et al., *Optical coherence tomography (OCT) reveals depth-resolved dynamics during functional brain activation.* Journal of Neuroscience Methods, 2009. **178**(1): p. 162-173.
19. Aguirre, A.D., et al., *Depth-resolved imaging of functional activation in the rat cerebral cortex using optical coherence tomography.* Optics Letters, 2006. **31**(23): p. 3459-3461.
20. Hope, J., et al., *Extracting morphometric information from rat sciatic nerve using optical coherence tomography.* Journal of Biomedical Optics, 2018. **23**(11).
21. Nam, A.S., et al., *Wide-Field Functional Microscopy of Peripheral Nerve Injury and Regeneration.* Scientific Reports, 2018. **8**(1): p. 14004.
22. Merkle, C.W., et al., *Dynamic Contrast Optical Coherence Tomography reveals laminar microvascular hemodynamics in the mouse neocortex in vivo.* NeuroImage, 2019. **202**: p. 116067.
23. Baran, U. and R.K. Wang, *Review of optical coherence tomography based angiography in neuroscience.* Neurophotonics, 2016. **3**(1): p. 010902-010902.
24. Liu, X. and J.U. Kang, *Depth-resolved blood oxygen saturation assessment using spectroscopic common-path Fourier domain optical coherence tomography.* IEEE transactions on bio-medical engineering, 2010. **57**(10): p. 2572-2575.

25. Yeh, Y.-J., et al., *Optical coherence tomography for cross-sectional imaging of neural activity.* Neurophotonics, 2015. **2**(3): p. 035001-035001.
26. Duncan, I.D. and A.B. Radcliff, *Inherited and acquired disorders of myelin: The underlying myelin pathology.* Exp Neurol, 2016. **283**(Pt B): p. 452-75.
27. de Campos Vidal, B., et al., *Anisotropic properties of the myelin sheath.* Acta Histochemica, 1980. **66**(1): p. 32-39.
28. Badreddine, A.H., *Optical tracking of nerve activity using intrinsic changes in birefringence*. 2017, Boston University.
29. Troiani, F., K. Nikolic, and T.G. Constandinou, *Simulating optical coherence tomography for observing nerve activity: A finite difference time domain bi-dimensional model.* PLOS ONE, 2018. **13**(7): p. e0200392.
30. Goodwin, M., et al., *Quantifying birefringence in the bovine model of early osteoarthritis using polarisation-sensitive optical coherence tomography and mechanical indentation.* Scientific Reports, 2018. **8**(1): p. 8568.
31. Chapman, C.A.R., et al., *Electrode fabrication and interface optimization for imaging of evoked peripheral nervous system activity with electrical impedance tomography (EIT).* Journal of Neural Engineering, 2019. **16**(1): p. 016001.
32. Hope, J., et al., *Extracting impedance changes from a frequency multiplexed signal during neural activity in sciatic nerve of rat: preliminary study in vitro.* Physiological Measurement, 2019. **40**(3): p. 034006.
33. Hope, J., et al., *Optimal frequency range for electrical impedance tomography of neural activity in peripheral nerve*, in *9th International IEEE EMBS Conference on Neural Engineering*. 2019: San Fransisco.
34. Fouchard, A., et al., *Functional monitoring of peripheral nerves from electrical impedance measurements.* Journal of Physiology-Paris, 2017.
35. Kassinopoulos, M., et al., *Correlation of the derivative as a robust estimator of scatterer size in optical coherence tomography (OCT).* Biomedical Optics Express, 2017. **8**(3): p. 1598-1606.
36. Kartakoullis, A., E. Bousi, and C. Pitris, *Scatterer size-based analysis of optical coherence tomography images using spectral estimation techniques.* Optics express, 2010. **18**(9): p. 9181-9191.
37. Hope, J., et al., *Increasing signal amplitude in electrical impedance tomography of neural activity using a parallel resistor inductor capacitor (RLC) circuit.* Journal of Neural Engineering, 2019.
38. Cohen, L.B., R.D. Keynes, and D. Landowne, *Changes in light scattering that accompany the action potential in squid giant axons: potential-dependent components.* The Journal of Physiology, 1972. **224**(3): p. 701-725.
39. Uttam, S. and Y. Liu, *Fourier phase in Fourier-domain optical coherence tomography.* Journal of the Optical Society of America. A, Optics, image science, and vision, 2015. **32**(12): p. 2286-2306.
40. Jacques, S.L., *Optical properties of biological tissues: a review.* Physics in Medicine and Biology, 2013. **58**(11): p. R37-R61.
41. Saltsberger, S., I. Steinberg, and I. Gannot, *Multilayer Mie scattering model for investigation of intracellular structural changes in the nucleolus and cytoplasm.* International Journal of Optics, 2012. **2012**.
42. Jonghwan Lee and David A. Boas (2012) "Frequency-Domain Measurement of Neuronal Activity using Dynamic Optical Coherence Tomography" 34th Annual Int Conf IEEE EMBS, San Diego, CA, USA.

**Supplementary Material**

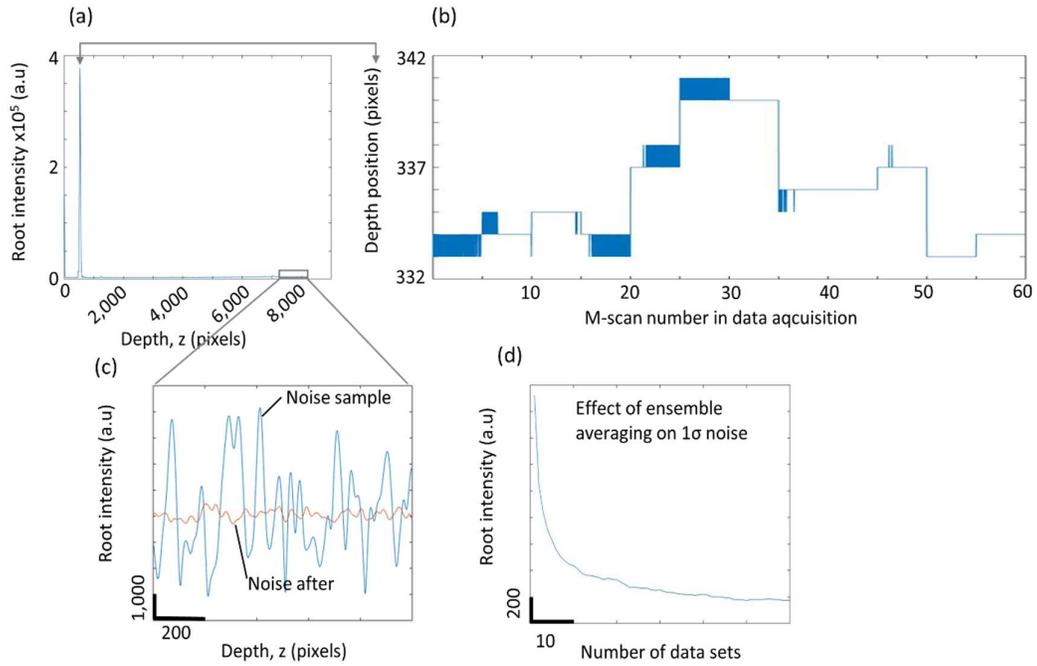

Figure S1: A-scan acquired on the mirror sample showing a single peak which corresponds to the mirror surface (a). Variation in the depth position of the surface peak within each M-scan and across the 60 M-scans of acquired data on the mirror (b). A noise sample taken from the deepest 1,000 pixels of an A-scan prior to and after ensemble averaging (c), and the relationship between 1 sigma noise and number of M-scans ensemble averaged (d).

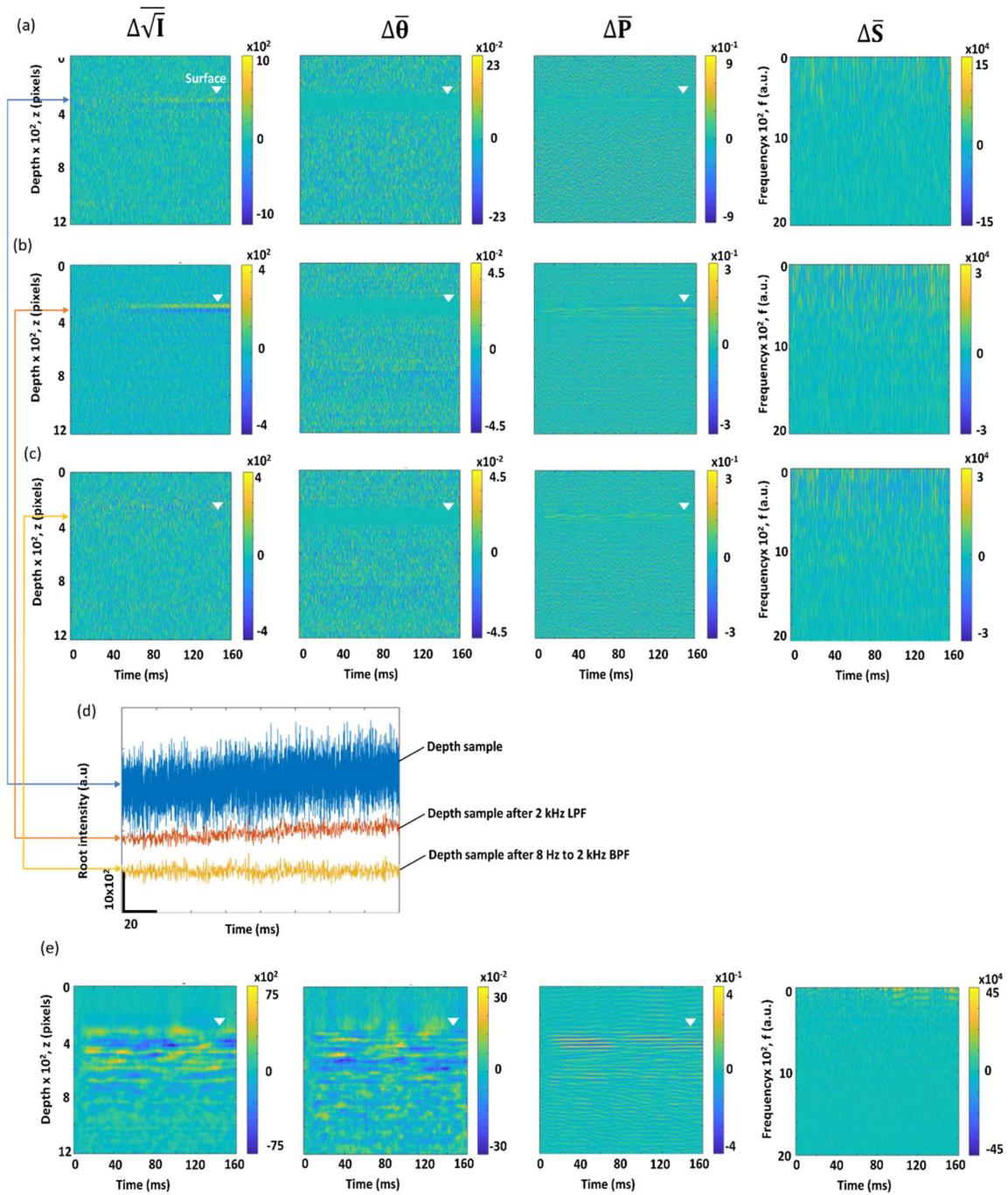

Figure S2: Intensity, phase retardation, phase, and spectra arrays from mirror before (a) and after (b and c) frequency filtering in the time dimension (d), and microspheres (e) after filtering. Bottom tips of the white triangles indicate the surface.

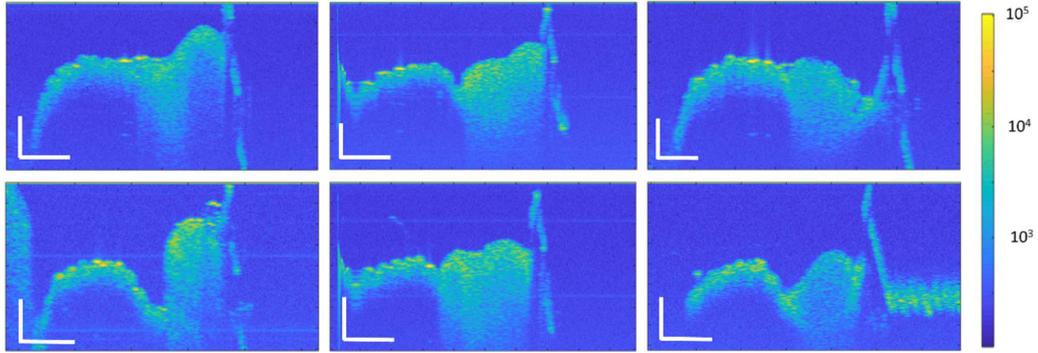

Figure S3: B-scans from several subjects of nerves in the nerve holder showing variation in size and shape, and little distortion of the nerve from mechanical loading. Vertical and horizontal scale bars in all images are 1,000 and 50 pixels, respectively. Vertical pixel size is 0.44 µm in nerve tissue, Width pixel size is 20 µm. Colour axis is log10 of intensity.